\documentclass[10pt,conference]{IEEEtran}
\IEEEoverridecommandlockouts

\usepackage{cite}
\usepackage{amsmath,amssymb,amsfonts}
\usepackage{algorithmic}
\usepackage{graphicx}
\usepackage{textcomp}
\usepackage{xcolor}
\usepackage{url}
\usepackage[hidelinks]{hyperref}
\def\BibTeX{{\rm B\kern-.05em{\sc i\kern-.025em b}\kern-.08emT\kern-.1667em\lower.7ex\hbox{E}\kern-.125emX}}

\begin{document}

\title{Using CognitIDE to  Capture Developers' Cognitive Load  via Physiological Activity During Everyday Software Development Tasks

    \thanks{The work of Fabian Stolp is funded by the Hasso Plattner Institute Research School on Data Science and Engineering. Charlotte Brandebusemeyer's work is funded by the SAP - HPI Research Program.
    *Both authors contributed equally.
    }
}

\makeatletter 
\newcommand{\linebreakand}{
  \end{@IEEEauthorhalign}
  \hfill\mbox{}\par
  \mbox{}\hfill\begin{@IEEEauthorhalign}
}
\makeatother

\author{
\IEEEauthorblockN{Fabian Stolp*}
\IEEEauthorblockA{\textit{Digital Health - Connected Healthcare} \\
\textit{Hasso Plattner Institute, University of Potsdam}\\
Potsdam, Germany \\
Fabian.Stolp@hpi.de}
\and
\IEEEauthorblockN{Charlotte Brandebusemeyer*}
\IEEEauthorblockA{\textit{Digital Health - Connected Healthcare} \\
\textit{Hasso Plattner Institute, University of Potsdam}\\
Potsdam, Germany \\
Char.Brandebusemeyer@hpi.de}

\linebreakand
\IEEEauthorblockN{Franziska Hradilak}
\IEEEauthorblockA{\textit{Digital Health - Connected Healthcare} \\
\textit{Hasso Plattner Institute, University of Potsdam} \\
Potsdam, Germany \\
Franziska.Hradilak@student.hpi.de}
\and
\IEEEauthorblockN{Lara Kursawe}
\IEEEauthorblockA{\textit{Digital Health - Connected Healthcare} \\
\textit{Hasso Plattner Institute, University of Potsdam} \\
Potsdam, Germany \\
Lara.Kursawe@student.hpi.de}

\linebreakand
\IEEEauthorblockN{Magnus Menger}
\IEEEauthorblockA{\textit{Digital Health - Connected Healthcare} \\
\textit{Hasso Plattner Institute, University of Potsdam}\\
Potsdam, Germany \\
Magnus.Menger@student.hpi.de}
\and
\IEEEauthorblockN{Franz Sauerwald}
\IEEEauthorblockA{\textit{Digital Health - Connected Healthcare} \\
\textit{Hasso Plattner Institute, University of Potsdam}\\
Potsdam, Germany \\
Franz.Sauerwald@student.hpi.de}

\linebreakand
\IEEEauthorblockN{Bert Arnrich}
\IEEEauthorblockA{\textit{Digital Health - Connected Healthcare} \\
\textit{Hasso Plattner Institute, University of Potsdam}\\
Potsdam, Germany \\
Bert.Arnrich@hpi.de}
}

\maketitle
\begin{abstract}
Integrated development environments (IDE) support developers in a variety of tasks. 
Unobtrusively capturing developers' cognitive load while working on different programming tasks could help optimize developers' work experience, increase their productivity, and positively impact code quality.

In this paper, we propose a study in which the \emph{IntelliJ}-based IDE plugin \emph{CognitIDE} is used to collect, map, and visualize software developers' physiological activity data while they are working on various software development tasks.
In a feasibility study, participants completed four simulated everyday working tasks of software developers -- coding, debugging, code documentation, and email writing -- based on Java open source code in the IDE whilst their physiological activity was recorded.
Between the tasks, the participants' perceived workload was assessed.
Feasibility testing showed that \emph{CognitIDE} could successfully be used for data collection sessions of one hour, which was the most extended duration tested and was well-perceived by those working with it.
Furthermore, the recorded physiological activity indicated higher cognitive load during working tasks compared to baseline recordings. 
This suggests that cognitive load can be assessed, mapped to code positions, visualized, and discussed with participants in such study setups with \emph{CognitIDE}.
These promising results indicate the usefulness of the plugin for diverse study workflows in a natural IDE environment.
\end{abstract}


\section{Introduction}
To improve software development efficiency, choosing a developer-centered approach is beneficial. 
Developers are content and more productive when their work can be completed without experiencing any friction \cite{brown_developer_2023}.
Potential points of friction have so far been assessed with log metrics, surveys, and diaries \cite{brown_developer_2023} and with source code metrics \cite{abbad-andaloussi_relationship_2023}.
These methods, however, only capture the developers' subjective experiences (surveys) or potential problems in the code regarding its quality (code metrics).
Having a tool that can detect perceived difficulties in sections of the source code objectively and in real-time would be beneficial in assessing cognitive load-inducing parts of the code and code quality.
Such a tool is the open source \emph{CognitIDE}\footnote{\url{https://github.com/HPI-CH/CognitIDE}} \cite{stolp_cognitide_2024}. 

Many software development tasks are tackled using the tools provided by Integrated Development Environments (IDE)s.
With the help of the IDE plugin \emph{CognitIDE}, it is possible to gather physiological data from body sensors and correlate them to problematic sections in the code and IDE usage that induce physiological reactions.

Building upon the introduction of our plugin in a paper by Stolp et al. \cite{stolp_cognitide_2024}, in this paper, we would like to highlight the usefulness of the plugin for empirical software engineering studies in which developers interact with source code.
User interface and developer experience studies that aim to analyze interactions with IDEs, software development activities, coding behavior, code quality, emotions, cognitive load, attention, and many more aspects in the broad context of developer -- source code / IDE interaction are supported by our plugin.
In the exemplary study setup and procedure we propose in this paper, we use these functionalities to let the participants complete typical software development tasks and interact naturally with the IDE while their physiological activity is recorded.
An automated workflow guides the participants through a pre-questionnaire, a relaxation video, baseline recordings, (programming) tasks with questionnaires appearing between the individual tasks, final baseline recordings, and a post-questionnaire without the experimenters needing to intervene.
The study procedure was reviewed and accepted by the University of Potsdam ethics committee.

In the following sections, we will provide background information on physiological recordings during IDE use and on the influence of cognitive load on software development productivity.
Additionally, we will describe the proposed study in more detail and share initial insights gathered from feasibility runs we conducted.

\begin{figure*}[!htb]
    \centering
    \includegraphics[width=0.75\linewidth]{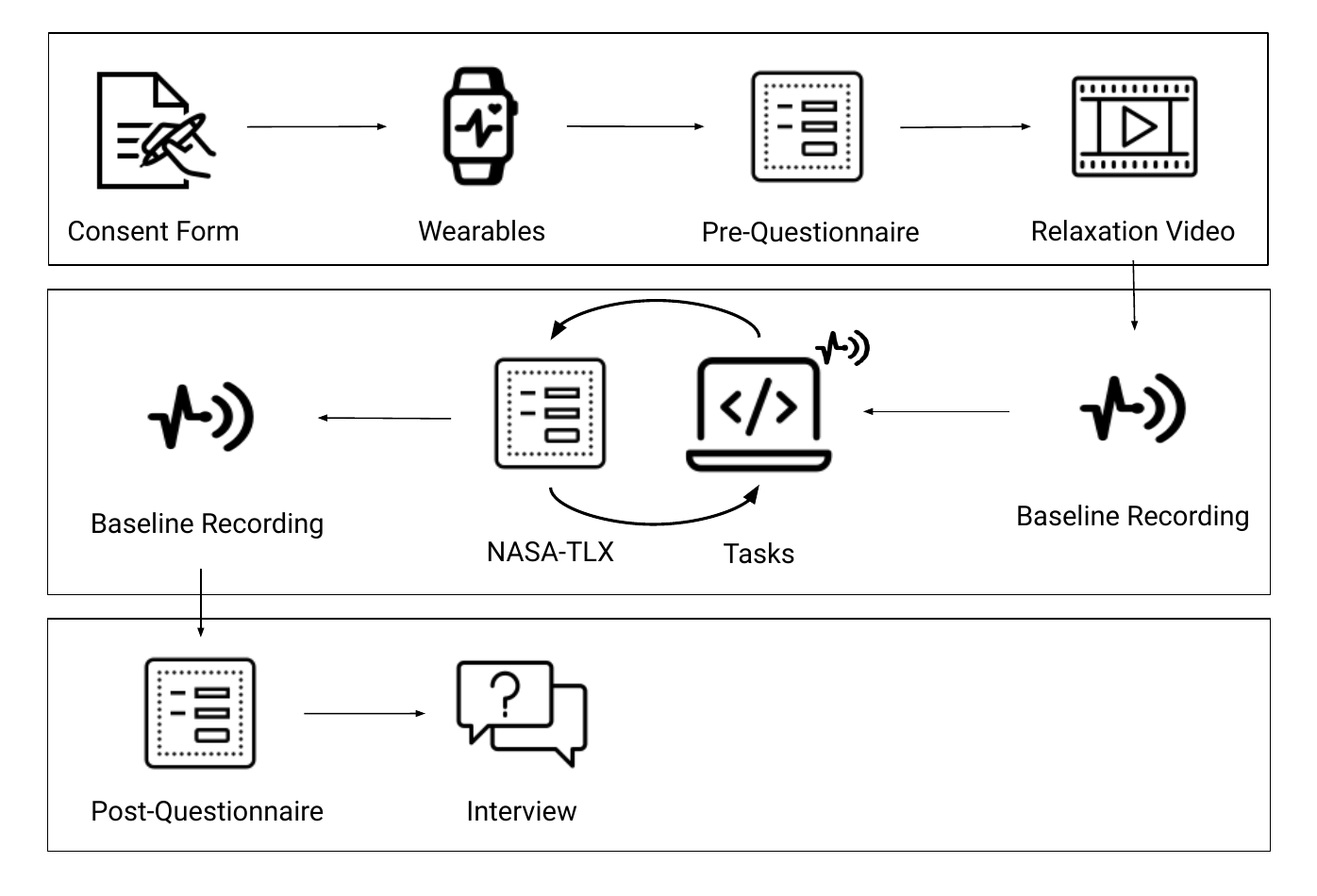}
    \caption{Overview of the study procedure. 
    As a first step, the participants were informed about the study and signed a consent form.
    After all wearables were attached and calibrated, \emph{CognitIDE} handles the study procedure automatically following the predefined workflow until after the post-questionnaire step.
    The workflow includes filling in questionnaires, relaxation videos, baseline recordings, and working on predefined software development tasks.
    After that, participants were interviewed by one of the study managers regarding their experiences during the experiment.\protect\footnotemark[11]}
    \label{fig:study-procedure}
\end{figure*}

\section{Background and related work}
Below we provide further information on research regarding physiological measurements during IDE use and cognitive load in software development.

\subsection{IDE use and physiological measurements}
IDEs support software developers during various tasks and their use is deeply integrated into developers' workdays.
While the developers' working experiences are coming more into focus in companies and research \cite{noda_devex_2023}, also interest in collecting physiological measurements during IDE use has increased \cite{weber_brain_2021}.

One of many motivations to assess physiological activity during IDE use is to support software developers' well-being.
The burnout rate among software developers worldwide is high \cite{jetbrains_state_2023}. 
By measuring developers' physiological activity and thereby deducing cognitive load during their working tasks in the IDE, early signs of burnout could be noticed \cite{pihlaja_occupational_2022}. 
Supplying feedback and charts to the user can improve awareness of their mental well-being and productivity-related information and metrics. 
Recommendations concerning breaks from work or efficient working task scheduling can be made.

The ability to map physiological data to concrete source code elements or IDE user interface components supports additional research directions:
Research on IDEs and further development of their usability can benefit from knowing where people are looking.
Through this information and potential further physiological indicators, one can comprehend which processes during IDE use induce cognitive load and therefore which elements are especially difficult to understand.
For example, concerning modern advances in generative artificial intelligence (GenAI) tools, one could investigate changes in the IDE user interface (UI) due to the integration of source code suggestions from large language models.
Another possible research question is why some software development tasks are especially challenging and what makes them demanding.
The proposed study design in this paper tackles the latter research question.

While working on such questions by conducting a study for assessing cognitive load through objective, continuous physiological measurements under controlled conditions \cite{stolp_assessing_2023}, we noticed that it would be highly beneficial to integrate stimulus or task presentation and body sensor data collection within IDEs to conduct such studies in less controlled real-world settings.

Tools such as \emph{iTrace} \cite{shaffer_itrace_2015}, \emph{Vitalse} \cite{roy_vitalse_2020}, and \emph{CognIDE} \cite{vieira_cognide_2020} that support collecting such data were described in the literature. 
However, none of these tools provided the flexibility we needed in terms of supported and simultaneously usable body sensors, code visualization capabilities, as well as task and questionnaire configurability and presentation.

Thus, we envisioned \emph{CognitIDE}. 
The plugin combines \emph{IntelliJ}-based IDEs with \emph{Lab Streaming Layer} (LSL).
LSL is "an open-source networked middleware ecosystem to stream, receive, synchronize, and record neural, physiological, and behavioral data streams acquired from diverse sensor hardware."\footnote{\url{https://labstreaminglayer.org/}}.

By integrating the LSL Java bindings\footnote{\url{https://github.com/labstreaminglayer/liblsl-Java}} into the \emph{CognitIDE} plugin, the feature of collecting body sensor data from various sources during everyday work tasks and settings was integrated into \emph{IntelliJ}-based IDEs. 
A list of supported devices can be found on the LSL website\footnote{\url{https://labstreaminglayer.readthedocs.io/info/supported_devices.html}}.
Using the sensor data that is collected by starting an LSL recording through \emph{CognitIDE}, the plugin calculates and leverages gaze positions to map additional sensor data to user interface components and code symbols that were looked at. 
For example, a person's heart, electrodermal, or brain activity can be linked with where the person looks in the code.
This mapping can then be used to highlight code based on the person's physiological activity.
Stronger highlighting can be configured to indicate higher physiological activity, which can be associated with a stress response due to heightened cognitive load while interacting with a specific potentially problematic code section.
Highlighting can occur on code symbol level but also on less granular levels.
Potential issues with code quality, which induce cognitive load and reduce productivity, can be detected this way.

In-depth technical details on the plugin are provided in the initial \emph{CognitIDE} publication \cite{stolp_cognitide_2024}. Information on setup and usage is provided in the corresponding repository\footnote{\url{https://github.com/HPI-CH/CognitIDE}}.
A minimal setup only requires a running installation of an IntelliJ-based IDE for Windows and a computer mouse, as \emph{CognitIDE} supports simulating gaze with the mouse pointer position.

When creating this tool, we strived to ensure high flexibility for different study designs: 
(1) building on top of the \emph{IntelliJ} platform -- supporting several different IDEs and programming languages, 
(2) integrating LSL -- supporting a wide variety of body sensors, and
(3) making the highlighting highly configurable by the user through scripting -- supporting code highlighting based on user-defined scripts. 
To facilitate further and more diverse study setups, the functionalities of the plugin were extended from those described in the initial \emph{CognitIDE} publication \cite{stolp_cognitide_2024}.
Among other features, the plugin now also supports automated workflows, code editing, and more configurable questionnaires.
Automated workflows enable researchers to define an order in which tasks, questionnaires, and other elements are shown to potential study participants within the IDE. 
This facilitates studies in which participants are supposed to work through tasks independently without being interrupted by study managers.
The features mentioned above are used in the study proposed in this paper.

\subsection{Influence of cognitive load on software development productivity}
Cognitive load can be defined as the amount of mental effort needed to carry out a task \cite{sweller_cognitive_2011}. 
Complex code structures and little documentation lead to increased cognitive load and, therefore, to more time and effort to complete a task successfully \cite{abbad-andaloussi_relationship_2023, fakhoury_effect_2018}.
The overall productivity of the developer is reduced.
Identifying the working tasks of a software developer that create high cognitive load is a first step to then detect which aspects make the tasks challenging.

So far, largely source code metrics are used to evaluate code quality \cite{abbad-andaloussi_relationship_2023, pantiuchina_improving_2018}. 
The majority of these metrics, however, could not be associated with cognitive load \cite{abbad-andaloussi_relationship_2023}.
In a systematic literature review on the relationship of source code metrics and cognitive load, Abbad-Andaloussi \cite{abbad-andaloussi_relationship_2023} concluded that it is difficult to have a metric that considers a developer's cognitive load whilst working on fragments of code in different classes, methods, etc.
He proposes future research to find a metric that captures the quality of task-related code only, to validate the relationship between source code metrics and cognitive load empirically, and to conduct experiments with larger source code sections.
The IDE plugin \emph{CognitIDE} enables researchers to tackle the above-proposed research suggestions.
Utilizing physiological data to indicate cognitive load and making it visible in the source code by highlighting sections that induced cognitive load allows the user to pinpoint problematic sections in the code. 
Insights from these sections can lead to the production of code that induces a lower cognitive load and, therefore, increases the code quality and productivity of the developer.

\section{Feasibility study}
\label{sec:feasibility_study}
\begin{figure*}[!htb]
    \centering
    \includegraphics[width=1\linewidth]{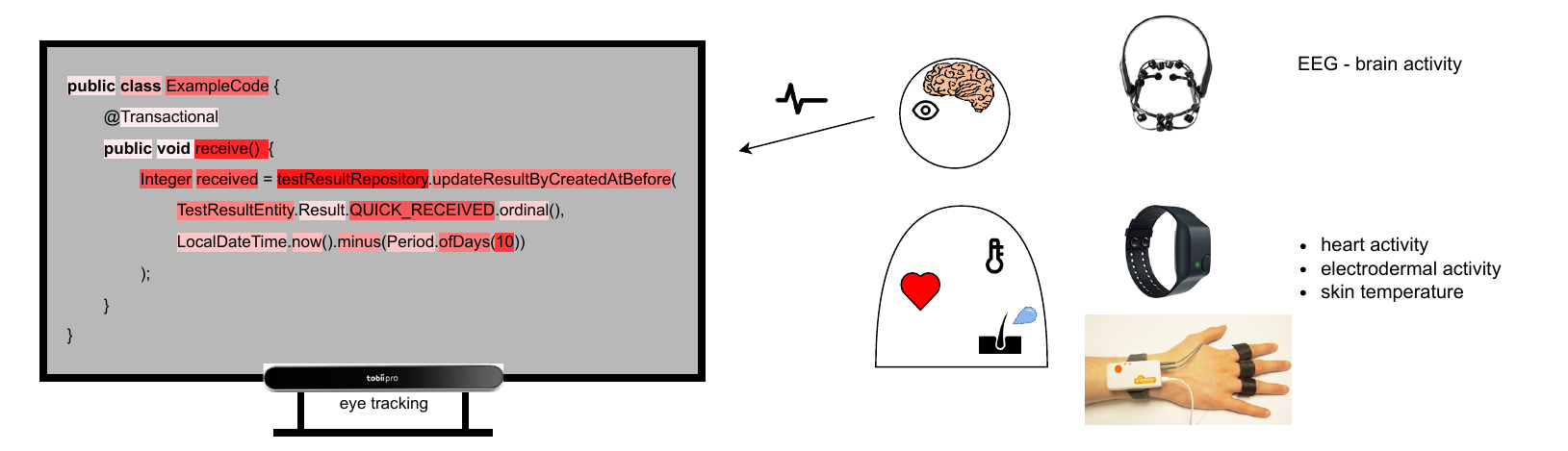}
    \caption{The mapping of a person's physiological activity onto an example source code is depicted. With the help of the eye tracker and \emph{CognitIDE}, it is possible to localize where the person is looking and color the source code according to heart activity, electrodermal activity, skin temperature, or brain activity.\protect\footnotemark[11]}
    \label{fig:study_setup}
\end{figure*}
An empirical feasibility study with four participants (two students, two research assistants, all male, age range: 23-29) was conducted to test our study design, the functionalities of \emph{CognitIDE}, the plugin's use for conducting studies on cognitive load, and to evaluate the user experience. \autoref{fig:study-procedure} provides a high-level overview of the study procedure.
The participants completed four simulated everyday working tasks of a software developer (coding, debugging, code documentation, and email writing) while interacting naturally with the IDE. 
The programming tasks are based on real-world Java open source code from the Corona Warn-App\footnote{\url{https://github.com/corona-warn-app}} -- an application widely used in Germany during the COVID-19 pandemic.
A total of 19 Java files with a total of 1402 lines of code were used.
Email writing was included as a task in the IDE to compare physiological activity during programming language interaction with that during natural language interaction.
Between the tasks, the widely used NASA Task Load Index (NASA-TLX) \cite{hancock_development_1988} to assess the subjective perceived workload of each task was displayed in the IDE and was filled in by the participants.
During the whole experiment, the physiological activity of the participants was recorded via body sensors. 
Eye movement\footnote{\url{https://www.tobii.com/products/eye-trackers/screen-based/tobii-pro-spark}} and brain\footnote{\url{https://www.emotiv.com/products/epoc-x}}, heart\footnote{\label{shimmer}\url{https://shimmersensing.com/product/shimmer3-gsr-unit/}}\footnote{\label{empatica}\url{https://empatica.com/en-eu/research/e4/}}, and electrodermal activity\footref{shimmer}\footref{empatica} were recorded to be later able to map and visualize the physiological activity in the IDE. 
Thereby, potentially problematic code sections that may be indicative of cognitive load can be highlighted.
\autoref{fig:study_setup} provides an overview of the setup.

At the end of the study, a brief evaluation in the form of an interview was conducted on the participant's experience during the study and the interactions with the plugin.

\section{Results}
\subsection{CognitIDE plugin}
The plugin functioned smoothly. All features worked as intended.
Initial concerns regarding the plugin's performance while recording a high amount of sensor data were not justified. 
No performance issues arose.
The recorded data did not exceed 300 MB for each participant after approximately one hour of recording.

During the interviews at the end of the study, the participants gave very positive feedback on the \emph{CognitIDE} plugin.
The predefined, automated workflow, i.e. the implemented study procedure, was highlighted as a major advantage by the participants.
This way, study conductors do not have to interrupt the study flow by explaining tasks or making adjustments.

\subsection{Measurement of cognitive load}
The analysis of the physiological data indicates higher physiological activity during the programming tasks compared to the baseline recording.
Pupil dilation increased, and electrodermal activity, as seen in a higher amount of skin conductance responses (SCRs) and SCR rise time, was higher.
Regarding brain activity, changes in alpha (7-13 Hz) and theta (4-7 Hz) frequency bands (de)synchronization could be observed during the tasks compared to the baseline, which stands in relation to cognitive load \cite{antonenko_using_2010}.
Results regarding heart activity were inconclusive due to sensor data that was too noisy.
In sum, the physiological activity changed while the participants were working on the tasks compared to the baseline recording, indicating the measurement of cognitive load.

\subsection{Visualization in the CognitIDE plugin}
The code sections were highlighted according to gaze duration at the end of the study and shown to the participants. 
Viewing time acted as a placeholder for more complex highlightings during the feasibility study.
The participants affirmed that the highlighting matched where they were looking the longest. 
Therefore, the mapping of physiological activity onto source code was perceived as accurate.

\section{Threats to validity}
While there are various categories of validity, similar studies mostly discuss construct, external, and internal validity \cite{wyrich_40_2023}. 
We will assess our study design based on these categories.

\subsection{Construct validity}
We base our study on prior research that showed the connection between physiological measurements and cognitive load. 
However, cognitive load is not the only aspect that influences physiological measurements; other bodily processes unrelated to information processing, e.g., emotional stress, also affect them.
To address this, we record a variety of modalities to better distinguish cognitive load indicators from others.

While we use questionnaires that have been used in previous studies, such as the NASA-TLX questionnaire, different participants might interpret the requested evaluation differently.
However, we provide all participants with precisely the same information and explanations.

In this study, we have a limited number of tasks that are not representative of all the potential tasks software developers face.
While we cannot claim that we provide a general evaluation of cognitive load-inducing tasks in software development through this study, we included tasks often necessary during software development and integrated them into a realistic code base and development environment.

\subsection{External validity}
The feasibility study described in this paper included students and research assistants instead of professional software developers.
While this helped test the study setup, we expect to see differences between this study population and professional software developers included in the upcoming study.
Related to that, although differences in physiological reactions between baseline and tasks could already be observed, a larger sample size is needed to reliably identify which specific tasks induce higher cognitive load than others.

We designed the study to be realistic regarding IDE use. However, additional factors, such as interruptions, tasks like meetings, and task durations of more than one hour, are outside the scope of this study and thus make the setting different from everyday work. 
Additionally, wearing body sensors like the used EEG headset is uncommon in natural work settings.

Last but not least, developers might be used to different development environments than IntelliJ-based IDEs.
However, the focus of the study tasks is not on using special IDE functions but on using basic functionalities that work similarly in IDEs to write, read, and alter code.

\subsection{Internal validity}
There can be individual differences in cognitive load responses.
Therefore, we take a holistic mixed-methods approach by considering perceived cognitive load from questionnaires and objectively measured physiological reactions.

While the participants were alone in the study room, they knew that the study managers could see them through a mirrored window. 
Feeling observed through sensors or people can potentially alter behavior.
In addition, it is possible that the wearables were noted more prominently at the beginning of the study because of the unusual feeling or at the end of the study after wearing them for a while.
Tasks were shown randomly between participants to mitigate these and similar effects on the recorded data.

\section{Future work}
Since positive feedback was given to the interaction with the plugin, physiological activity related to cognitive load could be measured in our study setup and the mapping of the physiological activity onto source code was successful, a study with the same setup will be conducted.
The upcoming study will include more participants who will be professional software developers for a more realistic real-life evaluation of the plugin-developer interplay.
Minor adjustments to the plugin and the study setup will be made.
In addition to gathering novel insights regarding the cognitive load connected to software development tasks, we hope to demonstrate the value of \emph{CognitIDE} as a research tool for empirical software engineering studies with wearables.

\section{Conclusion}
In this paper, we proposed a study examining cognitive load differences between software development tasks often worked on in IDEs. 
We described how the open source tool \emph{CognitIDE} can be used to execute this study and provided initial insights from feasibility runs.
The \emph{IntelliJ}-based IDE plugin enables us to examine the developer-source code interaction objectively. 
The developer's physiological activity can be visualized in the source code they worked on. 
Thereby, problematic sections that induce cognitive load, decrease productivity, and point toward code quality issues can be detected.
Identifying issues paves the way to improving developer experience and code quality.
Our exemplary study workflow was well perceived by the participants, cognitive load during everyday software development tasks could be measured with the body sensors, and the mapping of the physiological activity onto source code was perceived as accurate and precise.
Therefore, the additional functions of the plugin enable researchers and practitioners to conduct studies with diverse study workflows in a natural IDE environment.
Based on these promising findings from the feasibility runs, we will carry out the planned study in the future.

\section*{Acknowledgements}
We want to thank Amin Kanan and Stefan Spangenberg for their help in extending \emph{CognitIDE} and preparing and executing the test runs.
We additionally thank Dr. Ralf Teusner and Prof. Dr. Tobias Schimmer for adding ideas to the proposed study design.

\footnotetext[11]{Some icons in the figure are from \url{https://icons8.com/}.}

\bibliography{references}
\bibliographystyle{IEEEtran}

\end{document}